\documentclass[Journal]{IEEEtran}

\usepackage{graphicx, epstopdf}
\usepackage{setspace}
\usepackage{acronym}
\usepackage{amssymb}
\usepackage[cmex10]{amsmath}
\usepackage{commath}
\usepackage{mathrsfs}
\usepackage{mathtools}
\usepackage{ifthen}
\usepackage{textcomp}
\usepackage{float}
\usepackage[tight,footnotesize]{subfigure}
\usepackage{cite}
\usepackage{bm}
\usepackage{url}

\usepackage{multirow}
\usepackage{array}
\newcolumntype{L}[1]{>{\raggedright\let\newline\\\arraybackslash\hspace{0pt}}m{#1}}
\newcolumntype{C}[1]{>{\centering\let\newline\\\arraybackslash\hspace{0pt}}m{#1}}
\newcolumntype{R}[1]{>{\raggedleft\let\newline\\\arraybackslash\hspace{0pt}}m{#1}}

\IEEEoverridecommandlockouts

\begin{document}

\title{Molecular Communications: Channel Model and Physical Layer Techniques}

\author{\authorblockN{Weisi Guo,~\IEEEmembership{Member,~IEEE,} Taufiq Asyhari,~\IEEEmembership{Member,~IEEE,} Nariman Farsad,~\IEEEmembership{Member,~IEEE,} \\H. Birkan Yilmaz,~\IEEEmembership{Member,~IEEE,} Bin Li,~\IEEEmembership{Member,~IEEE,} Andrew Eckford,~\IEEEmembership{Senior~Member,~IEEE,} and Chan-Byoung Chae,~\IEEEmembership{Senior~Member,~IEEE}}

\thanks{W. Guo is with the University of Warwick, UK. T. Asyhari is with the University of Bradford, UK. N. Farsad and A. Eckford are with York University, Canada. B. Li is with Beijing University of Posts and Telecommunications, China. H. B. Yilmaz and C.-B. Chae (Corresponding Author) are with Yonsei University, Korea.}}

\maketitle

\begin{abstract}
This article examines recent research in molecular communications from a telecommunications system design perspective.  In particular, it focuses on channel models and state-of-the-art physical layer techniques.  The goal is to provide a foundation for higher layer research and motivation for research and development of functional prototypes.  In the first part of the article, we focus on the channel and noise model, comparing molecular and radio-wave pathloss formulae. In the second part, the article examines, equipped with the appropriate channel knowledge, the design of appropriate modulation and error correction coding schemes. The third reviews transmitter and receiver side signal processing methods that suppress inter-symbol-interference. Taken together, the three parts present a series of physical layer techniques that are necessary to producing reliable and practical molecular communications.
\end{abstract}

\begin{keywords}
Molecular communication, error correction coding, modulation, signal processing, channel modeling.
\end{keywords}

\section{Introduction}
Molecular communication utilizes chemical molecules as an alternative carrier for information.  Tiny devices (having dimensions in microns or less) may find molecular communications to be an efficient and low-cost alternative to high frequency electromagnetic based systems. Indeed, this is perhaps why molecular communications is prevalent in nature, both at the inter-organism and at the inter-cell scales. Being able to communicate at the nano-scale is important in unlocking the futuristic possibilities in nanotechnology. For some time now, the fundamental building blocks have been available for nanomachines. The transformative applications of nanotechnology have one feature in common: they involve not just single devices working independently, but swarms of devices working in concert (i.e., a network of nano-machines \cite{Akylidiz12}). Thus before the promise of nanotechnology can be fully achieved, the problem that remains to be solved is communication among nanodevices. The economic potential is enormous: the present market size is at \$100 billion, growing at a projected 12.3\% per year to \$200 billion in 2020.

The present capabilities of molecular communication are rather primitive. Indeed, the world's first molecular communication system capable of sending a text message has a volume of roughly 100 cm$^{3}$.  Akin to spectral efficiency, the system has a \emph{chemical efficiency} of 0.3 bits/s/chemical over a free-space distance of a few meters~\cite{Farsad13PLOS}. To have a significant impact, these systems will need to shrink and increase their data rates by two orders of magnitude.  The data rate increase can be achieved by: designing tailor-made modulation and error correction coding schemes~\cite{Shih13JSAC}, and increasing the number of orthogonal chemical compounds \cite{Chae13} or carrier frequencies \cite{Guo15CL} used to carry information.  Drawing on the current experience of designing radio-wave communication systems, a three-fold increase in data rate can be yielded with signal processing techniques.  A similar gain is expected through molecular communications.

\begin{figure*}[t]
    \centering
    \includegraphics[width=1.80\columnwidth]{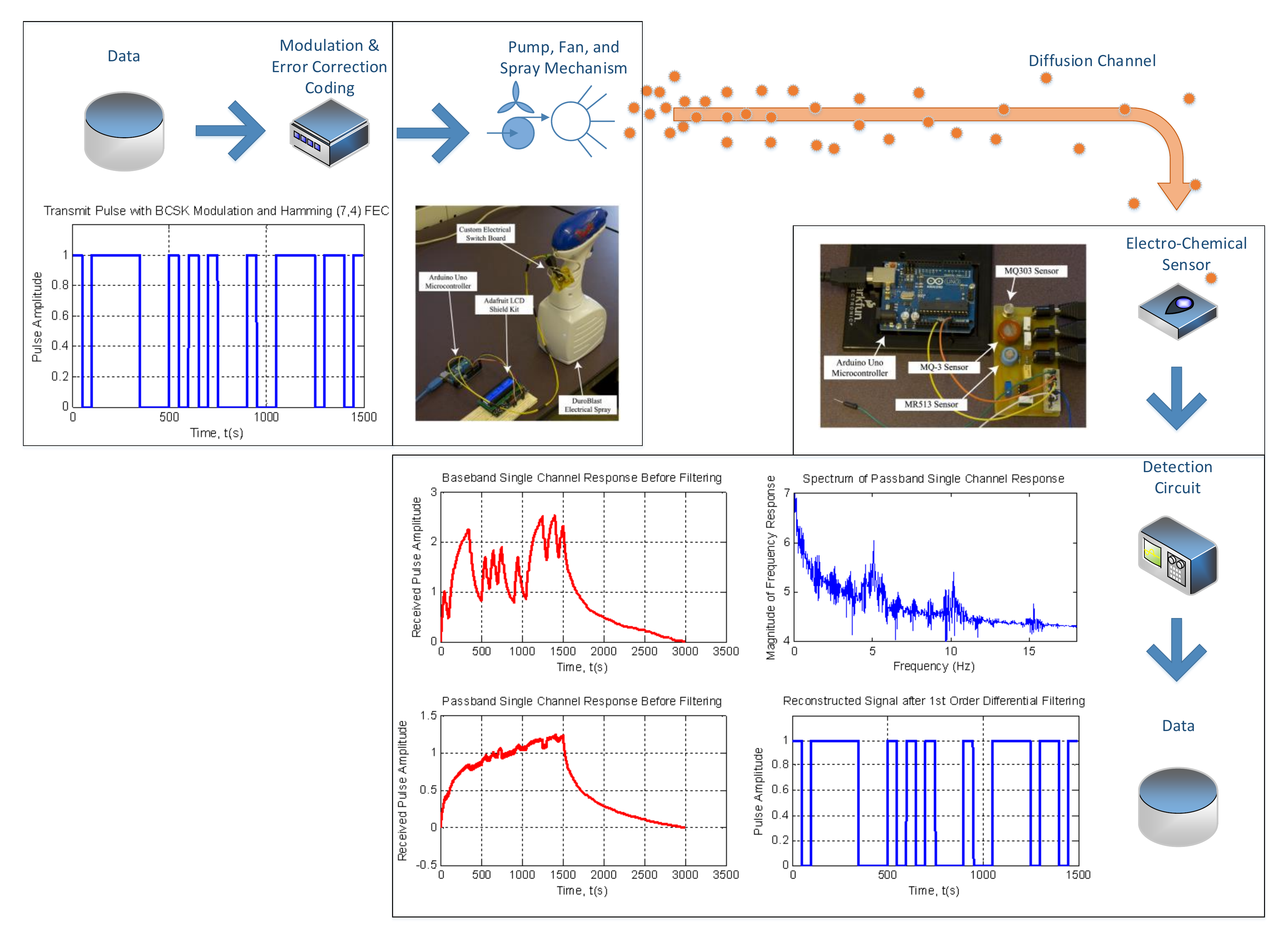}
    \caption{Illustration of molecular communication system physical layer components, hardware implementation, and the state of the signal at each stage.}
    \label{System}
\end{figure*}

This article reviews recent state-of-the-art-physical layer (PHY) techniques that go towards increasing chemical efficiency.  Figure~\ref{System} offers an illustration of a molecular communication system physical layer design, hardware implementation and the associated signal at each stage. The goal of this article is to provide a foundation for higher layer research, i.e., relay, full duplex, and multiple-input multiple-output (MIMO), as well as motivation for further research and development.

\section{Channel, Pathloss, and Noise Models}

\subsection{Statistical Channel Models}

\subsubsection{Hitting Distribution}
For a basic random walk process, a common equation of interest is the probability density function, $h(d,t)$, of emitted particles from a point source hitting a point in space. Indeed the equation determines the communication aspects of the channel for a given distance $d$, time $t$, and diffusivity $D$.  The time dependent particle distribution function differs in different environments.  The hitting distribution is given as: $h(d,t) = \frac{1}{(4\pi D t)^{a/2}}\exp(-d^{2}/4Dt)$, where $a$ is a constant. This model assumes that molecules that hit the receiver can undergo a random walk and hit the receiver again at a later point.

\subsubsection{Absorbing Receiver}
In nature, most of the receptor types remove the information molecules from the environment once they arrive at the receiver. This absorption or capture process is modeled by first hitting processes where molecules are removed from the environment after hitting the absorbing receiver. In a 1-dimensional (1-D) environment, the rate of capture is given by $h_{c}(d,t) = \frac{d}{(4\pi D t^{3})^{1/2}}\exp(-d^{2}/4Dt)$ for distance $d$ and time $t$. The probability of hitting to an absorbing receiver until time $t$, is $F_{c}(d,t) = \text{erfc} \left( d / \sqrt{4Dt\,} \right)$.  Similarly, the fraction of hitting molecules to a perfectly absorbing spherical receiver of radius $r_r$ in a 3-D environment is \cite{yilmaz2014ThreeDChannelCF} $h_{c}(d,t) = \frac{r_r}{d+r_r} \, \frac{d}{(4\pi D t^{3})^{1/2}}\exp(-d^{2}/4Dt)$, which is similar to the 1-D case. Note that there is a positive probability of not hitting the absorbing boundary for a diffusing particle in a 3-D environment even when time goes to infinity.

For signal processing and system analysis, it is also useful to know the system transfer function in the complex $s$-domain.  Assuming a linear time invariant (LTI) system, the transfer function for $h_{c}(d,t)$ is given by \cite{Wang14INFOCOM}: \begin{equation}
    \label{Transfer_Function}
    H_{c}(s) = \mathcal{L} [h_{c}(d,t)]= \exp\bigg(-d\sqrt{\frac{s}{D}}\bigg).
\end{equation}

\subsection{Pathloss Models: Free-Space and Complex Environments}

Pathloss models are concerned with the rate of energy decay over distance. Such models are vital for calculating the link budget of a telecommunication channel.  We first review the pathloss model in an unbounded free-space environment for both electromagnetic (EM) waves and molecular diffusion.  Assuming the EM waves radiate from a point source, the pathloss at distance $d$ is: $(4\pi fd/c)^{-2}$, where $f$ is the frequency and $c$ is the speed of light in a vacuum.  The time of arrival of the wave is $\tau = d/c$.  For molecules that diffuse from a point source in free-space, we propose two measures of loss:
\begin{enumerate}
  \item \emph{Peak molecular response:} The maximum pulse amplitude is detected at $\tau=d^{2}/6D$, which yields a peak hitting amplitude of \cite{Llaster14}: $\propto D/d^{-3}$, and a peak absorbing amplitude of \cite{yilmaz2014ThreeDChannelCF}: $\propto\frac{r_r \,D}{(r_r+d) \,d^2}$.
  \item \emph{Total molecular response:} This can be derived by counting the total number of molecules hitting the receiver over all time \cite{Llaster14, Guo15TMBMC} to be: $(4\pi Dd)^{-1}$, and absorbed by the receiver to be \cite{yilmaz2014ThreeDChannelCF}: $r_r/(r_r+d)$.
\end{enumerate}

Therefore, when comparing EM wave and molecular pathloss in free-space, one finds that:
\begin{enumerate}
  \item \textbf{EM:} the pathloss is $\propto d^{-2}$ and time-of-arrival is $\propto d$;
  \item \textbf{Molecular:} the total energy pathloss is $\propto d^{-1}$, the peak energy pathloss is $\propto d^{-3}$, and time-of-arrival is $\propto d^{2}$;
\end{enumerate}
That is to say, the peak molecular energy decays rapidly with distance, but the total molecular energy decays at a lower rate than EM-waves, but the time-of-arrival is quadratic with distance, which is significantly larger than EM-wave propagation.

In terms of more complex environments, a comparison was recently made for pipe, knife-edge, aperture, and mesh channels \cite{Guo15TMBMC}. It was found that unlike EM-wave signals, molecular propagation has the following propagation advantages:
\begin{itemize}
  \item \textbf{No Diffraction Loss:} does not suffer from diffraction loss when in the shadow of objects;
  \item \textbf{Not Obstacle Size Sensitive:} is not restricted by cut-off frequency in pipe, aperture, and mesh environments;
  \item \textbf{Shortest Path Dominates:} molecular pathloss only depends on the shortest path distance between the transmitter and receiver.
\end{itemize}
These propagation advantages for molecular communications can be exploited in certain communication scenarios listed in \cite{Guo15TMBMC}.  The caveats to molecular communications are numerous.  Besides the obvious slow nature of diffusion and its dependency on environmental changes such as temperature and flow, molecular communication is also at risk of degradation either from chemical reactions or bacteria consuming bio-molecules.

\subsection{Additive Noise and ISI Models}

In terms of additive noise, the vast majority of molecular communications research has been modeling based, utilizing classic diffusion equations with an additive noise component.  In a real pulse modulated molecular communication system, noise can arise for many reasons and it is difficult to take all of them into account or argue which may dominate.  Examples of noise include, but are not limited to: inconsistent mechanical pulse emission, atmospheric contamination, physical disturbances, and receiver design \cite{Pierobon11}.  Recent research has yielded two noise models for different scenarios:
\begin{enumerate}
  \item \textbf{High-Drift Velocity:} In \cite{Farsad14}, based on the hardware prototype introduced in \cite{Farsad13PLOS}, it was found that the Gaussian distributed additive noise was a reasonable approximation in a high velocity turbulent flow channel, which is found in application areas such as gas pipes.
  \item \textbf{Zero Drift:} As for a random walk diffusion channel with no induced drift velocity, the noise arises from the random arrival of molecules.  As shown in \cite{Kuran10}, for $n$ emitted molecules, the number of molecules arriving at the receiver is Binomial distributed $\sim \mathcal{B}(n, h_{c}(d,t))$.  Such a channel is common in low pressure environments such as caves and tunnels.
\end{enumerate}

The inter-symbol-interference (ISI) distribution has only been analyzed for a binary concentration modulated scheme in \cite{Kuran10}.  Given that the number of molecules arriving at the receiver is Binomial distributed $\sim \mathcal{B}(n, h_{c}(d,t))$, and that $n$ is large and the probability $h_{c}$ is neither close to 1 nor 0, the ISI can be approximated as a sum of random variables, each following a normal distribution.  The underlying assumptions of this widely used model are: (i) the ISI is dominated only by the previous symbol, and (ii) the transmission distance is neither too close ($h_{c} \rightarrow 1$) nor too far ($h_{c} \rightarrow 0$). This distribution has been extensively used in subsequent work for capacity analysis and the evaluation of error correction code performances \cite{Leeson14EL}.  This remains an open area for more complicated $M$-ary modulation schemes or those that have a high symbol rate.

\section{Modulation Schemes}

\subsection{Baseband Modulation}

\subsubsection{Basic Schemes}
For a fixed transmission distance of $d$, one can observe that there are essentially two main properties of the channel to modulate: the number of transmitter molecules $M$; and the pulse delay time $T_{k}$, such that the channel response is $h_{k}(t-T_{k})$.  For an input of binary symbols $a_{k} \in \mathcal{A} = \{0,1\}$, $k = 0, 1, \ldots, \infty$, the output of the baseband pulse modulator is $M_{k}$.  Existing pulse modulation can be summarized as being one of the following \cite{Guo15CL}:
\begin{itemize}
  \item Amplitude / Concentration Shift Keying (ASK or CSK), where the information is modulated into different levels of $M_{k}$, i.e., Binary CSK (BCSK): $M_{k} \in \mathcal{M} = \{0,M\}$.
  \item Frequency shift keying (FSK), where a sinusoidal pulse of a variable frequency $f$ is emitted $M_{k}(f_{k}) = M\sin(2\pi f_{k}t)$, i.e., Binary FSK: $f_{k} \in \mathcal{F} = \{0,f\}$.
  \item Pulse Position Modulation (PPM), where the information is modulated into the bit delay time $T$, i.e., Binary PPM: $T_{k} \in \mathcal{T} = \{0,T\}$.
\end{itemize}
In addition, the chemical composition can be used to encode information, as it is common in nature, which is known as Molecule Shift Keying (MoSK) \cite{Chae13}.  However, while MoSK is common in nature, the engineering complexity of synthesizing and detecting even a small number of precise chemical compositions is complex and expensive.

\subsubsection{ISI Reducing Schemes}
Molecular concentration shift keying (MCSK) is proposed in~\cite{arjmandi2013diffusionBN}, in which two molecule types are used for consecutive slots at the transmitter for suppressing the interference. The transmitter uses molecule type $A_1$ in odd time slots and molecule type $A_2$ in even time slots. The difference with MoSK is that the keying depends on the amplitude not the type of the molecule. On the receiver side, according to slot number only the $A_1-$ or $A_2-$ type molecules are counted. Since the molecule types are different in the two subsequent time slots, interference due to the previous symbol is drastically reduced.  Molecular Transition Shift Keying (MTSK) was also proposed as an energy-efficient ISI-reducing modulation technique for high data rates~\cite{tepekule2014energyEI}. MTSK employs multiple molecule types. The bit-0s in the sequence are encoded by the absence of the messenger molecules. The bit-1s are encoded by using two different types of molecules, where the choice of the molecule type depends on the value of the next symbol. If the next symbol is a transition from bit-1 to bit-0 (i.e., the final bit-1 for consecutive bit-1s) then the molecule type is changed.
\begin{figure}[t]
    \centering
    \includegraphics[width=1.0\linewidth]{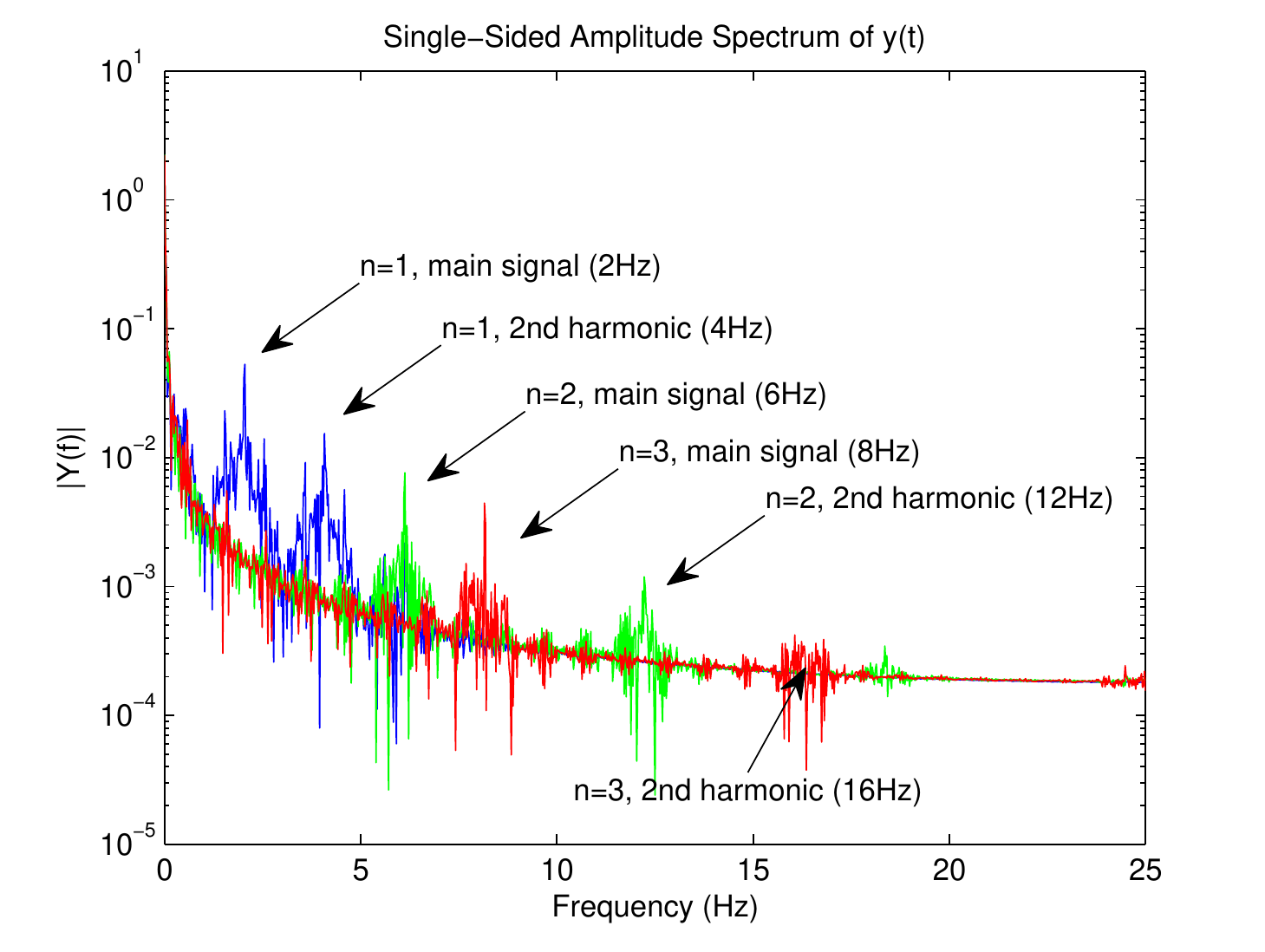}
    \caption{Frequency response of $N=3$ molecular signals multiplexed over a single diffusion channel with BCSK baseband modulation.}
    \label{FrequencyResponse}
\end{figure}

\subsection{Passband Modulation}

The aforementioned modulation schemes can be regarded as \emph{baseband modulation}, whereby only a limited data rate can be achieved.  The concept of a carrier wave, such as that associated with electromagnetic wave communications, is absent in molecular communications.  A recent breakthrough detailed in \cite{Guo15CL} shows that by adding an initial longitudinal vibration at the transmitter at frequency $f_{c}$, the spectra of the received molecules will exhibit this frequency component.  This is achieved by creating an oscillating transmitter that varies the distance of transmission for each data stream, such that multiple data streams can be carried on different \emph{carrier frequencies} (i.e., $f_{c,1}, f_{c,2}, ... f_{c,N}$). Figure~\ref{FrequencyResponse} shows the magnitude of the received frequency response $|Y(f)|$ of $N=3$ independent data channels.  The data streams are transmitted via a common diffusion channel using the same chemical to a common receiver.  It can be seen that the \emph{baseband signal} is at 0 normalized frequency.  The first harmonic of each signal can be seen distinctively, which is at the respective carrier frequencies.  The distinctive frequency components can be separated using bandpass filters.  Further research into this area will examine the capabilities regarding ISI suppression and the fundamental limits to multiplexing.

\section{Forward Error Correction Coding}

Forward-Error-Correction (FEC) coding is an important aspect of modern digital communications. In this section, we first examine how FEC codes can combat these two types of errors, namely: additive noise errors for CSK modulation schemes, and transposition errors for PPM modulation schemes; and we then compare the performance of different coding strategies.

\subsection{Additive Noise Errors}

In the presence of additive noise, the \emph{Hamming distance} is commonly used as a criteria for code word design.  Relatively simple block codes use added parity check bits to enable the correction of errors.  Hamming codes are described as $n = 2^{m}-1$ length codes, where $m$ is the number of parity check bits.  The minimum distance of this type of block code is 3, allowing the detection of 2 errors and the correction of 1 error. In order to develop low energy nano-communication systems, minimum energy FEC codes were considered in \cite{Leeson14EL}.  When considering a BCSK modulation scheme, codewords with a lower weight (more `0' bits) result in a lower energy consumption. Therefore, minimizing the average code weight is necessary to minimizing the energy expenditure.  According to the results in \cite{Leeson14EL}, this coding scheme is able to achieve significantly lower energy per bit (50\% reduction), but will need large codeword lengths to achieve a similar level of Bit-Error Rate (BER) performance.  More advanced codes such as Self-Orthogonal Convolution Codes (SOCCs) have also been developed recently and improve on Hamming codes to deliver a superior coding gain and lower energy expenditure.

\subsection{Bit Transposition Errors}

When block coding is considered, bit transposition may result into two sources of errors:
\begin{enumerate}
  \item \textbf{Intra-Codeword Errors} due to permutation of bits position within a codeword (crossovers within the same codeword), and
  \item \textbf{Inter-Codeword Interference} due to bits late/early arrival that affects other codewords (crossovers from the contiguous codewords).
\end{enumerate}

\subsubsection{Intra-Codeword Errors}
Consider encoding each of $M$ messages into $n$-bit block. A predominant coding technique is to select $M$ out of $2^n$ codewords that maximize the minimum pairwise Hamming distance. This approach is particularly relevant when bit distortion due to additive noises is the main source of errors. However, this may no longer be accurate when the dominant errors are due to bit transposition where instead of being distorted, bits exchange their positions. A key to deal with intra-codeword errors is to identify relevant attributes that are still preserved when bits within a codeword exchange positions. Hamming weight is a sensible choice that fulfills this requirement. The coding technique then follows by constructing a codebook that consists of words having distinct Hamming weights. We refer to the resulting code as a \emph{distinct Hamming-weight (DHW) code}. An example of DHW code for $n = 4$ and rate-$\frac{1}{2}$ is given by: $\mathcal{C}_{\rm dhw} = \{0000,\quad 1000, \quad 1100, \quad 1110 \}$.  The maximum coding rate of DHW code with block length $n$ is given by: $R_{\text{weight}} = \log_{2}(n+1)/n$.

\subsubsection{Inter-Codeword Errors}
Molecular coding (MoCo) distance function has been recently proposed as a new criterion for code design that tackles intra- and inter-codeword errors \cite{Shih13JSAC}. The MoCo distance of a pair of codewords $(\mathbf{x}_{i}, \mathbf{x}_j)$ is given by $d_{\rm MoCo} (\mathbf{x}_i, \mathbf{x}_j) \triangleq - \log \left( \Pr \{ \mathbf{x}_j | \mathbf{x}_i  \}\right)$, where $\Pr \{ \mathbf{x}_j | \mathbf{x}_i \}$ is the transition probability from $\mathbf{x}_i$ to $\mathbf{x}_j$ that is defined by the random walks of the molecules. The MoCo code construction is essentially finding codewords that maximize the minimum pairwise MoCo distance. For $n=4$ and rate-$\frac{1}{2}$, the MoCo code is given by: $\mathcal{C}_{\rm MoCo} = \{0000,\quad 1000, \quad 0010, \quad 1110 \}$. While addressing intra-codeword errors is clear from the MoCo distance, the least significant bit `0' in each codeword was believed to serve as a guard band that reduces inter-codeword interference. Unfortunately, the nature of exhaustive search for the best codebook and the lack of analytical expression of the error probability for good MoCo codes make it difficult to assess whether $\mathcal{C}_{\rm MoCo}$ will still be optimal for varying system and channel parameters.

\subsubsection{ISI-Free Code}

A well specified family of $(n,k,\ell)$ ISI-free codes was proposed in \cite{Shih13JSAC} to address both intra codeword errors and inter codeword interference. Herein $n$, $k$ and $\ell$ denote the block length, the message length (in bits) and the correctable crossover level, respectively. Table~\ref{table:ISI-free} provides an example of ISI-free codes when $(n,k,\ell) = (4,2,1)$. As observed from Table~\ref{table:ISI-free}, intra-codeword errors are mitigated using codewords with distinct Hamming weights, similarly to the DHW code. The inter-codeword interference is suppressed by assigning two codewords for each message such that for two contiguous codewords, the last $\ell$ bits of the first codeword are identical to the first $\ell$ bits of the second codeword. Encoding messages $01,10,11$ yields codewords $0001,1100, 0111$.

\begin{figure}[t]
    \centering
    \includegraphics[width=1.0\columnwidth]{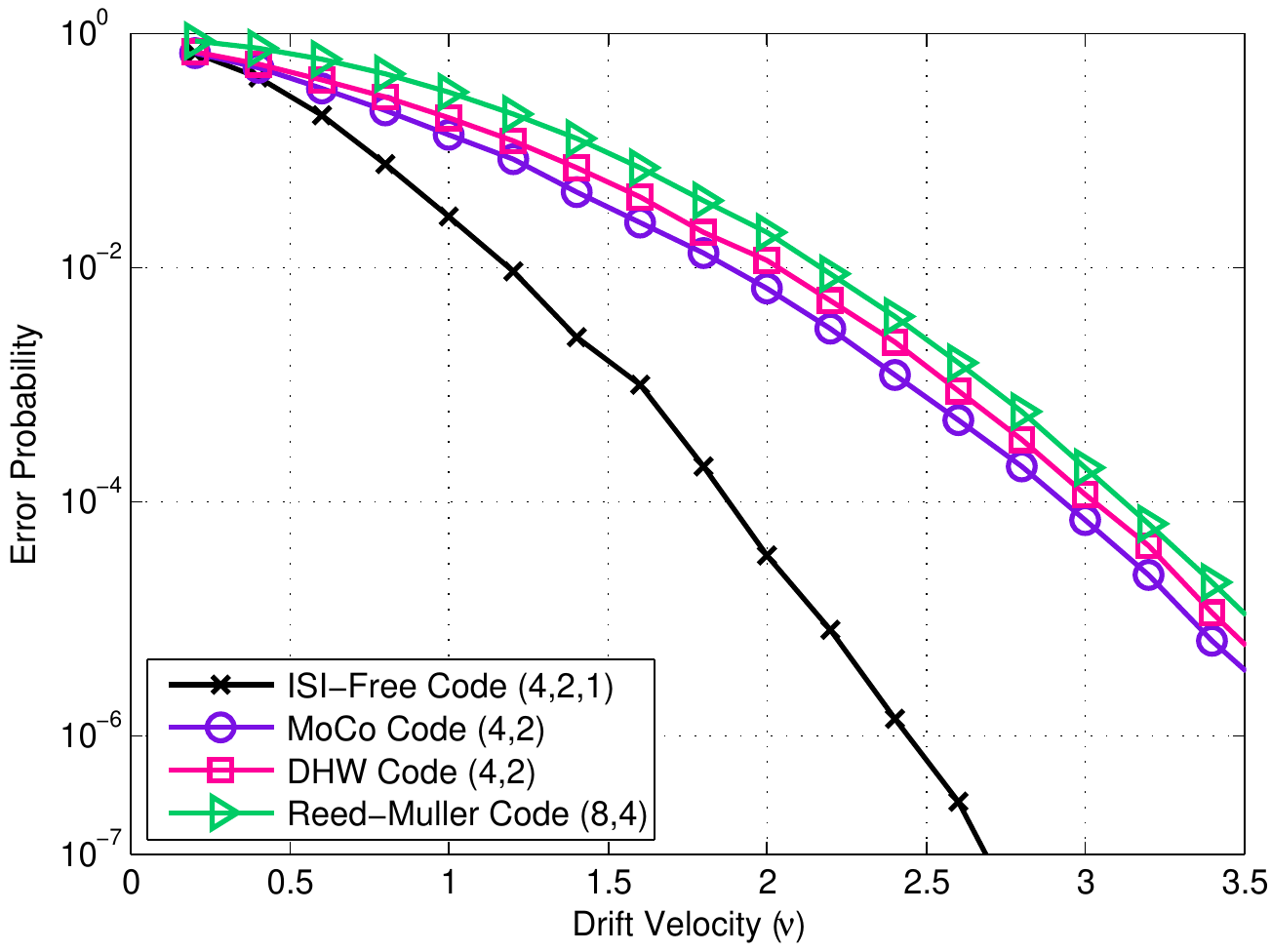}
    \caption{Error probability of various coding schemes as a function of drift velocity $v$ ($\mu m/s$) in a PPM system: $D = 1 \: \mu m^2/s$, $d = 1$ $\mu m$, and all the coding schemes have rate of $1/2$. }
    \label{FEC}
\end{figure}

\begin{table}
\caption{(4,2,1) ISI-free code with codewords assignment.}
%\begin{tabular}{| C{2.3cm} | C{2.5cm} | C{2.5cm} |}
\begin{center}
\begin{tabular}{| c| c | c |}
\hline
\multirow{3}{*}{Information Bits} & \multicolumn{2}{ c| }{Codeword if  the last bit of the}  \\
								  & \multicolumn{2}{ c| }{previous codeword is}  \\ \cline{2-3}
			        & ~~~ bit `0' 	~~~		      &  bit `1'		\\
\hline
00			&	0000						   &  1111 					\\ \hline
01			&	0001						   &  1000 					\\ \hline
10			&	0011						   &  1100					\\ \hline
11			& 	0111						   &  1110					\\
\hline
\end{tabular}
\end{center}
\label{table:ISI-free}
\end{table}

\subsection{Performance Comparison}
Figure~\ref{FEC} demonstrates the performance of different coding schemes over diffusion-based molecular communications. It can be observed that Reed-Muller code (8,4) designed based on Hamming distance is inferior to other codes. For a single codeword assignment for each message, the MoCo code is superior to the DHW code since the MoCo code has been suggested to address both intra-codeword errors and inter-codeword interference. The main drawback of the MoCo code is the lack of structured construction. Among all the presented codes in Fig.~\ref{FEC}, $(4,2,1)$ ISI-free code has the best error performance. This code is able to guarantee free transposition errors up to $\ell = 1$ level.

With the availability of structured construction and superior performance, ISI-free codes appear to be promising for molecular communications. However, further developments are necessary to address their inherent rate inefficiencies. In the above example, due to the assignment of two codewords for a single message, the coding rate above $\frac{n-1}{n}$ cannot be achieved for ISI-free codes. Some adjustments have been proposed in \cite{Shih13JSAC}, but the rate improvement is still limited. Furthermore, the inherent code construction in \cite{Shih13JSAC} assumes bits `0' and `1' being transmitted using two molecule types. This is inefficient, not only due to the cost of using different types of chemical substance, but also due to the two molecule types that alternatively would have allowed a higher multiplexing data rate. In order to overcome these shortcomings, a key research direction would be comprehensive understanding of the trade-offs between rate and error probability using information-theoretic approaches.

\section{Chemical Techniques for ISI Suppression}

\subsection{Molecular Degradation}

As mentioned previously, ISI arises from the long tail of the channel response. Therefore, it is logical to consider if chemical degradation can mitigate ISI. Molecular degradation is governed by ${E + S  \xrightleftharpoons[k_{-1}]{\,k_{1}\,} ES \xrightharpoonup{k_p} E + P}$, where $E$, $S$, $ES$, and $P$ denote the enzyme, substrate, enzyme-substrate compound, and the product, respectively. Reaction constants are denoted by $k_1$, $k_{-1}$, and $k_p$ \cite{heren2014effectOD_ARXIV}. Considering the nature of the molecular communication scenario simplifies the equation dynamics and leads to ${\frac{d [S]}{dt} = - \lambda [S]}$, where $[\cdot]$ denotes the concentration and $\lambda$ is the rate of degradation. Instead of $\lambda$, sometimes half-life of the molecules, $\Lambda_{1/2}$, is considered in the equations. Having a very high degradation rate may cause molecules to degrade so fast that they cannot arrive at the target receiver. Therefore, for each symbol duration there exists an optimal degradation rate.
\begin{figure}[t]
\centering
\includegraphics[width=1.0\columnwidth]{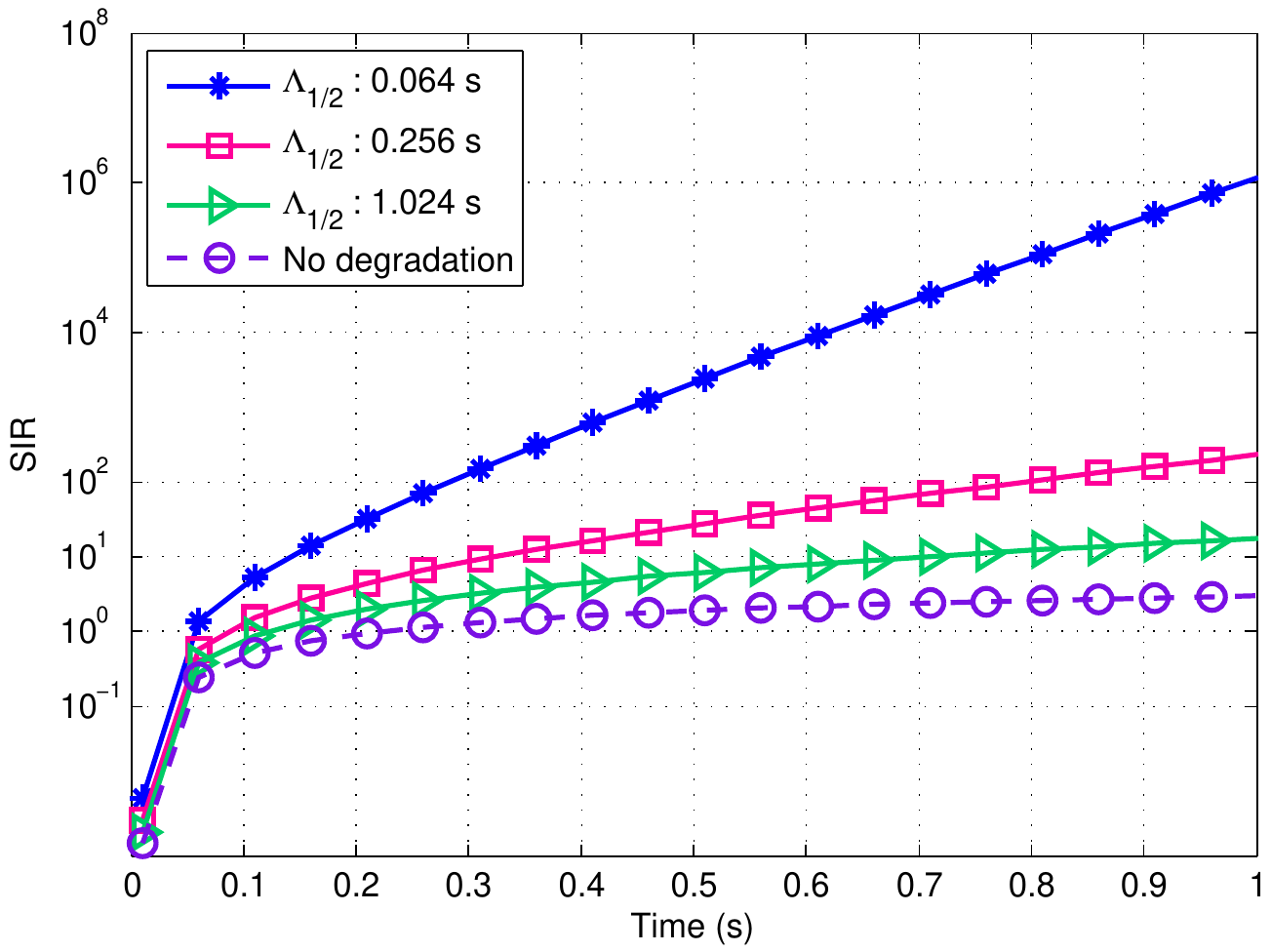}
\caption{SIR versus symbol duration for various $\Lambda_{1/2}$ (half-life) values. Faster degradation clears stray molecules and reduces the amount of ISI in the system. ($d = 4~\mu m$, $r_r=10~\mu m$, $D=79.4 ~\mu m^2/s$).
}
\label{fig:Rate_of_arrivals}
\end{figure}

When the molecular degradation is taken into account, channel characteristics differ. Heavy tail property of the arriving molecules changes and results in a decrease in ISI with the cost of requiring more molecules to be synthesized. The authors in~\cite{heren2014effectOD_ARXIV} analyzed the effect of degradation for a given symbol duration $t_s$. A good metric for observing the effect of symbol duration is the signal-to-interference ratio (SIR). SIR is basically the ratio of molecules that arrive in the first symbol duration to the ratio of ISI molecules and given as $ \frac{F_{c}^{\lambda}(d,t_s)} {F_{c}^{\lambda}(d,t\rightarrow \infty)-F_{c}^{\lambda}(d,t_s)}$.  Figure \ref{fig:Rate_of_arrivals} depicts SIR for various half-life values, where faster degradation clearly reduces the amount of stray molecules for a given molecular communication scenario. Note that the reduction in ISI comes with a cost of higher pathloss due to decomposed molecules.  This effect is left for future investigations.

\subsection{Transmitter Side Pulse Shaping with Chemical Reaction}

An alternative degradation method is outlined in \cite{Wang14INFOCOM}, which uses reactive pulses. Assuming that the random walk channel is an LTI system, as mentioned earlier, the channel transfer function is found in Eq.~\eqref{Transfer_Function}. It can be shown, using either the principle of channel inversion or designing a window function, that the necessary transmit pulse is of the form\footnote{The validity of this proof only exists for when the distance is small or the diffusivity is high ($x \ll \sqrt{D}$).}: $x(t) = \delta(t) - \frac{x}{\sqrt{D}}t^{-3/2}$, where the two parts of the equation can be interpreted as follows.  The $\delta(t)$ is the information bearing pulse.  The negative decay element $\propto -t^{-3/2}$ is a poison signal designed to chemically react with the information pulse and cancel the tail of the transient response at the receiver (positioned at distance $x$).  The constant of proportionality is $\frac{x}{\sqrt{D}}$, which means the channel state information at transmitter (CSIT) is needed.  Specifically, this is the transmission range $x$ and the channel's molecular diffusivity $D$.  While simulation results in \cite{Wang14INFOCOM} show that the system response using pulse shaping will yield fewer errors, how we emit the signal is not obvious (given its divergent and undefined nature at $t=0$).  This is a problem for future research to find practical pulse shapes.

\begin{figure}[t]
    \centering
    \includegraphics[width=1.0\linewidth]{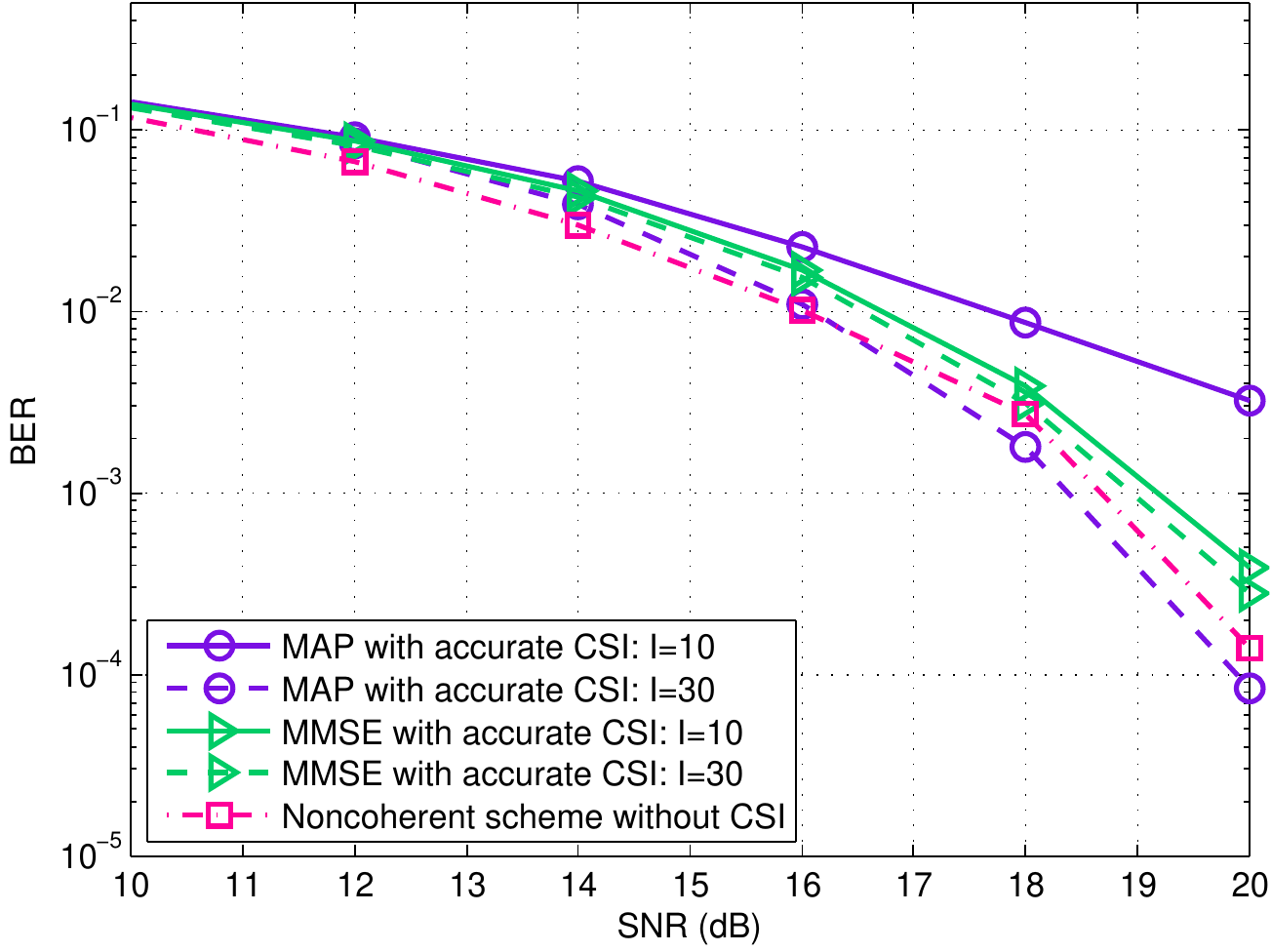}
    \caption{BER performance of coherent detection schemes (MAP and MMSE) and the non-coherent scheme under memory-length $I$.  Parameters used are: 3D diffusion channel, distance 2$m$, diffusivity 0.1$cm$$^{2}$$/s$, receiver absorption radius 2$cm$, and the symbol period is 4$s$.  }
    \label{BER_coherent_vs_noncoherent_different_I}
\end{figure}

\section{Receiver Side Signal Processing}

On the receiver side, equalizer filters can be used to eliminate the ISI effect. The received signal vector is normally expressed as $\mathbf{r=Ya+w}$, where $\mathbf{r}=[r_{0},r_{1},\ldots,r_{K}]^{T}$, $\mathbf{Y}$ is the circulant channel matrix constructed from $\mathbf{y}$, and $\mathbf{w}$ is noise vector.

For a coherent MAP scheme, the accurate estimation of unknown channel impulse response (CIR) should be acquired firstly. To reduce the complexity, a sequential estimation strategy will be used in practice. For the i.i.d. information source, the MAP scheme is also equal to the maximum likelihood (ML) method.  Another widely used sub-optimal detector is based on the MMSE criterion, which minimizes the covariance matrix of the detection errors.  Unlike the additive Gaussian noise, variance of the received signal is history dependent and equalizer tap coefficients must be adjusted according to previous detected bits for MMSE filters. Hence, the tap count affects the computational complexity. On the other hand, the computational complexity of the proposed decision feedback filter (DFF) is independent of the tap count. However, this requires more memory $I$ to achieve the same BER.

It is recognized that, in sharp contrast to wireless communications, there are two inherent challenges in molecular communications. First, the CIR cannot be easily estimated without a pilot channel, especially in the presence of channel disturbances due to air flow and temperature variations.  Even in the presence of a pilot channel, the coherence period of the channel is usually small, while the channel delay is large. Second, molecular communication is targeted towards nano-scale systems, which require a low computational complexity system to reduce hardware costs and energy expenditure.  The effectiveness of the coherent MAP or MMSE algorithms in eliminating ISI largely depends on both accurate CSI and complicated matrix or polynomial (likelihood) operations.  Based on the above considerations, coherent detectors are less attractive in the context of low-complexity and low-power nano-machines.

Difference-based signal detectors can be used as a simple non-coherent and low complexity detector.  In the hardware prototype found in \cite{Farsad13PLOS}, a tunable difference detector is used to compare the molecular concentration at two time steps.  In comparing the performance of the aforementioned non-coherent difference detector and the previous coherent detectors, we plot the BER as a function of the signal-to-noise-ratio (SNR).  The SNR is defined as the ratio of the peak signal power and the Gaussian noise due to stochastic molecule arrival \cite{Kuran10}. As shown in Fig.~\ref{BER_coherent_vs_noncoherent_different_I}, the BER performance of a non-coherent detector without CSI can exceed existing coherent and more complex MMSE and MAP schemes that use a memory-length of $I=10$.

\section{Conclusions}
This article has reviewed state-of-the-art physical layer techniques developed over the past few years. The article first focused on the advantages of molecular propagation over radio waves, which suffer from diffraction loss and cut-off frequencies. We then reviewed the design of appropriate modulation schemes, especially those that suppress ISI and add longitudinal carrier frequency components to create bandwidth. For error correction codes, two types were analyzed: (i) low energy Hamming codes when in the presence of additive noise, and (ii) distinctive Hamming weighting and positional distance codes in the presence of bit disposition errors.  Finally, the paper reviewed signal processing methods that suppress ISI, including chemical reactions and low-complexity non-coherent detection schemes. Taken together, the article presented a series of advanced physical layer techniques that are necessary to ensure reliable and practical molecular communications.

\bibliographystyle{IEEEtran}
\bibliography{IEEEabrv,CM_Ref}

\end{document}